\documentclass[useAMS,usenatbib]{mn2e}
\usepackage{amssymb}
\usepackage{graphicx}
\usepackage{mathrsfs}
\usepackage{bm}
\title[Relativistic mixing-layer model for FR\,I jets] {A relativistic
mixing-layer model for jets in low-luminosity radio galaxies} \author[Y. Wang et
al.]{Y. Wang$^{1}$\thanks{E-mail: wangyang@astro.soton.ac.uk},
C. R. Kaiser$^{1}$, R. Laing $^{2}$, P. Alexander$^{3}$, G. Pavlovski$^{1}$ and
C. Knigge$^{1}$\\$^{1}$School of Physics and Astronomy, University of
Southampton, University Road, Southampton SO17 1BJ\\$^{2}$European Southern
Observatory, Karl-Schwarzschild-Stra\ss e 2, D-85748 Garching-bei-M\"{u}nchen,
Germany\\$^{3}$Astrophysics Group, Cavendish Laboratory, University of
Cambridge, J J Thomson Avenue, Cambridge CB3 0HE}

\begin{document}

\pagerange{\pageref{firstpage}--\pageref{lastpage}} \pubyear{2009}

\maketitle

\label{firstpage}

\begin{abstract}
We present an analytical model for jets in Fanaroff \& Riley Class I (FR\,I)
radio galaxies, in which an initially laminar, relativistic flow is surrounded
by a shear layer.  We apply the appropriate conservation laws to constrain the
jet parameters, starting the model where the radio emission is observed to
brighten abruptly. We assume that the laminar flow fills the jet there and that
pressure balance with the surroundings is maintained from that point
outwards. Entrainment continuously injects new material into the jet and forms a
shear layer, which contains material from both the environment and the laminar
core. The shear layer expands rapidly with distance until finally the core
disappears, and all of the material is mixed into the shear layer. Beyond this
point, the shear layer expands in a cone and decelerates smoothly. We apply our
model to the well-observed FR\,I source 3C\,31 and show that there is a
self-consistent solution. We derive the jet power, together with the variations
of mass flux and and entrainment rate with distance from the nucleus. The
predicted variation of bulk velocity with distance in the outer parts of the
jets is in good agreement with model fits to VLA observations.  Our 
prediction for the shape of the laminar core can be tested with
higher-resolution imaging.
\end{abstract}


\section{Introduction}
\label{introduction}

Extragalactic radio sources were divided into two morphological classes by
\citet{fr74}. FR\,I sources have an edge-darkened structure, whereas FR\,II
sources are edge-brightened with prominent outer hot-spots. This classification
has proved to be extremely robust: the division between the classes depends
primarily on radio luminosity \citep{fr74}, with FR\,II sources being more
powerful, but also on the stellar luminosity of the host galaxy \citep{lo96}.
There are significant differences between the structures of the jets in the two
classes: those in FR\,I sources often flare close to the nucleus and have large
opening angles, whereas their equivalents in FR\,II sources are highly
collimated out to the hot-spots \citep{bridle84}. There is good evidence that
FR\,I jets are initially relativistic, but decelerate on kiloparsec scales,
whereas FR\,II jets remain relativistic until they terminate (e.g.\
\citealt{Laing93}).

The process of deceleration in FR\,I jets appears to be complex, and may involve
a transition to turbulent flow.  In addition, the sources have a wide range of
morphologies, ranging from well-defined lobes similar to those in FR\,II sources
to extended plumes or tails \citep*{parma96}.  For these reasons, attempts to
construct global models of the evolution of FR\,I sources, linking observable
quantities such as linear size and radio luminosity, have been less
straightforward than the equivalents for FR\,II sources (e.g.\
\citealt*{Scheuer74,ka97,kda97}), which assume that the jet flows are
essentially laminar.  Part of the motivation for the present study is therefore
to construct a simple model of FR\,I jets for use as input to global models.  We
consider twin-jet sources, which make up at least one half of the FR\,I
population \citep{parma96}, excluding wide-angle tail and fat-double sources
\citep{ol89,ow91}, whose jet properties differ significantly.

Over the last few years, detailed modelling of deep VLA observations of jets in
five FR\,I sources has allowed us to quantify their geometries, velocity
distributions, magnetic fields and emissivity distributions in three
dimensions. We refer in detail to the analysis of 3C\,31 by \citet[hereafter
LB02a]{lb02a}; observations and models of a further four sources have
subsequently been published \citep{cl04,canvin05,lcbh06}. A consistent picture
of FR\,I jet deceleration on kiloparsec scales has emerged from these
studies. The flow velocities are $\beta = v/c \approx$ 0.8 -- 0.9 where the jets first
brighten abruptly, typically at $\sim$1\,kpc from the nucleus.  The jets flare
and then recollimate, decelerating rapidly to speeds
of $\beta \approx$ 0.1 -- 0.4.  The best-fitting transverse velocity profiles
appear to be approximately self-similar. At least in the 4/5 cases where the
jets appear to be propagating in contact with the interstellar medium of the
host galaxy rather than inside radio lobes, they are roughly 30\% faster on-axis
than at their edges.  Nevertheless, an evolution of the velocity profiles with
distance from the nucleus is not excluded. In particular, the transverse
velocity variations are poorly constrained where the jets first brighten
abruptly and a top-hat profile would also be consistent with the observations in
these regions of all five sources.

In order to decelerate, a jet must entrain matter, either from stars within its
volume \citep{Phinney83,komi94} or by ingestion of the surrounding material at
its boundary, as originally suggested by \citet{baan80}, \citet{deyoung81} and
\citet{begelman82}. In the latter case, the transverse velocity profile almost
inevitably evolves with distance from the nucleus.

X-ray observations can be used to infer the temperature, density and pressure
profiles of the hot gas associated with the host galaxies of FR\,I radio
galaxies \citep[e.g.][]{hardcastle02,worrall03,hardcastle05}.  Together with the
velocity distributions derived from modelling of the radio emission, these can
be used in a conservation-law analysis \citep[hereafter B94]{bick94} to derive jet energy fluxes and the
variations of mass flux, pressure, internal density and entrainment rate with
distance from the nucleus \citep[hereafter LB02b]{lb02b}. Such an analysis is
quasi-one-dimensional and therefore adopts values for the flow variables (in
particular the velocity) averaged across the jet cross-section.  This is
reasonable if the velocity profiles have restricted ranges and do not evolve
significantly with distance down the jets, as is consistent with the
observations of 3C\,31 (LB02b). If FR\,I jets are in pressure equilibrium
with their surroundings after they recollimate, this analysis requires that a
significant overpressure drives the initial flaring.

An alternative approach, which would also be consistent with the observations,
is to postulate that the transverse velocity profiles evolve significantly as
the jets interact with the external medium.  The first approximation is then to
assume pressure equilibrium between the jet and its surroundings and to take
explicit account of the interaction between the jets and their surroundings
using a simple mixing-layer model. This is the subject of the present paper.
The key assumption is that there is a turbulent mixing layer between the jet and
its environment, produced by the interaction of the two components.  The mixing
layer grows both into the jet and into the environment, and the initially
laminar jet eventually becomes fully turbulent.  As in the quasi-one-dimensional
analysis of \citet{lb02b}, we use the relativistic formulation of the laws of
conservation of mass, momentum and energy given by B94.

We describe the geometry of the jet-layer model in Section~\ref{structure}. The
relativistic conservation laws are introduced in Section~\ref{laws}. We derive
and discuss the solutions for our model in Section~\ref{solutions}. In
Section~\ref{3c31}, we apply our model to observations of 3C\,31. We discuss the
effects of varying model parameters in Section~\ref{discussion} and summarize
our conclusions in Section~\ref{conclusion}.

\section{Structure of an FR\,I jet}
\label{structure}

\begin{figure*}
\includegraphics[width=0.8\textwidth]{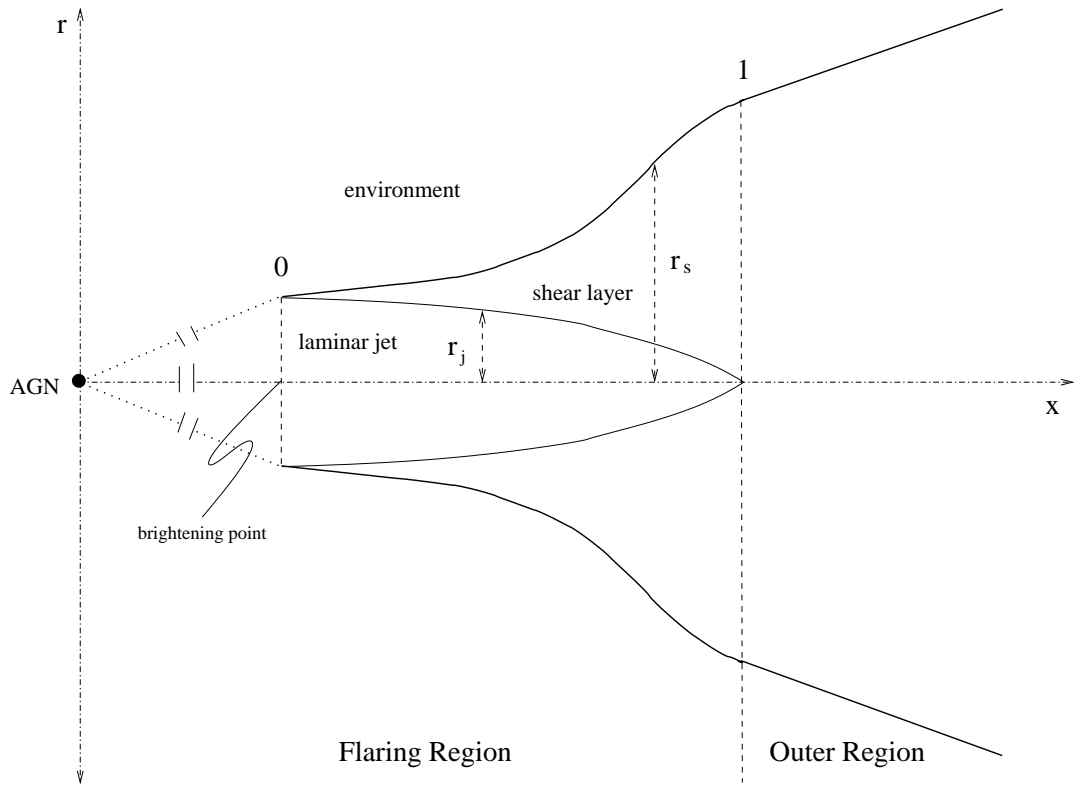}
\caption{A sketch of the principal features of our jet model (not to scale). For
  comparison with later figures, the brightening point in 3C\,31 is 1.1\,kpc
  from the nucleus and the transition between the flaring and outer regions is
  at 3.5\,kpc (see Fig.~\ref{flare}a)} \label{cartoon}
\end{figure*}

The basic structure of an FR\,I jet in our model is shown in
Figure~\ref{cartoon}. Following the definition given by LB02a, we divide the jet
into {\em flaring} and {\em outer} regions.\footnote{LB02a postulated the
existence of an additional conical inner region in the faint inner jets of
3C\,31, but observations of the better-resolved source NGC\,315 by
\citet{canvin05} are inconsistent with a constant expansion rate in the
corresponding part of the brighter jet. A continuously increasing expansion rate
is required in NGC\,315 and is equally consistent with the observations of
3C\,31 and other sources. A two-zone model is adequate to describe the geometry
in all cases.}  Close to the nucleus in the flaring region, the
outer isophotes have small, but increasing opening angles. Further out, they
spread rapidly and then recollimate. In the outer region, the expansion is
conical. The radio emission close to the base of the flaring region is usually
faint and it is always possible to identify a distance from the nucleus where
the jet brightens abruptly. We refer to this location as the {\em brightening
point}\footnote{This is also a change of terminology from LB02a, who refer to
the {\em flaring point}, and is done to emphasise that the location marks a
change in emissivity profile, not in geometry}.

We assume pressure equilibrium with the surroundings at all distances from the
nucleus and adopt the simplest possible prescription for velocity variations
following \citet{cr91}.  Wherever possible, we approximate the velocity of a
component of the flow by its spatially averaged value. We postulate that the
flow close to the axis of the flaring region is laminar, with a constant
relativistic bulk velocity $v_{j}$ and that this occupies the full width of
the jet at the brightening point, where interaction with the external medium
becomes significant for the first time. As a result of entrainment of external
material, a slower {\em shear layer} forms between the laminar jet and the
environment. The bulk velocity of the shear layer should vary continuously in the radial
($r$) direction from $v_j$ at its boundary with the laminar core to 0 at its
outer edge.  The models derived by LB02a show, however, that the ratio of edge
to centre velocity for the synchrotron-emitting material is $\approx$0.7
throughout the flaring and outer regions, so the velocity range within the
mixing layer is fairly narrow. Thus we assume that the steady-state flow in this layer has a constant bulk velocity
$v_{s}<v_{j}$. Material from both the environment and the laminar jet is 
continuously injected into the shear layer, the latter component supplying
energy and momentum as well as mass.  Integrated across the jet, the fraction of
slower material then increases with distance from the nucleus; this would be
interpreted as deceleration of the entire flow in fits to observations with poor
transverse resolution.

The laminar jet in the centre eventually vanishes, so no more energy or
momentum can be injected into the shear layer from the inside. Motivated by the
analysis of 3C\,31 (LB02a), we assume that this transition occurs precisely at
the end of the flaring region. This may not be general: modelling of other
sources suggests that the bulk of the jet deceleration occurs in the first part 
of the flaring region (e.g.\ NGC\,315; \citealt{canvin05}).  We assume that the
boundary of the shear layer in the outer region expands smoothly and more slowly
as the environmental pressure decreases. Entrainment from the environment into
the shear layer can still happen in the outer region, but this requires that the
velocity be allowed to vary along the jet (Section~\ref{cons-outer}). We assume
that there are no transverse velocity gradients.

The following convention is adopted throughout this paper: we use subscript 0
for quantities at the brightening point; 1 for quantities at the end of the
flaring region; $j$, $l$ and $e$ for all quantities related to the laminar jet,
shear layer and environment, respectively.  Detailed descriptions of the
parameters are given in Fig.~\ref{cartoon} and Table~\ref{parameter1}.

\section{Relativistic conservation laws}
\label{laws}

We model the structure of FR\,I jets using relativistic fluid mechanics,
applying the laws of conservation of mass, momentum and energy in the forms
given by B94. As in that reference, we use the relativistic enthalpy
$\omega=\rho c^{2}+\epsilon+p$ and the ratio $\mathscr{R}=\rho
c^{2}/(\epsilon+p)$ of rest-mass energy to non-relativistic enthalpy. Here,
$\epsilon$ is the internal energy density and $p$ is the pressure. In the laminar jet, we expect $\mathscr{R}_{j} \propto p^{-1/4}$ (B94). For
3C\,31, the external pressure drops by a factor $\approx 4$ from the brightening
point until the end of the flaring region. The approximation that
$\mathscr{R}_{j}$ is constant is therefore reasonable, and we adopt it in what
follows.

For an ideal gas, $\epsilon=p/(\Gamma-1)$, so $\mathscr{R}$ can be
written as:
\begin{equation}
\mathscr{R}=\frac{\Gamma-1}{\Gamma}\frac{\rho
c^{2}}{p}=\frac{\Gamma-1}{\Gamma}\frac{\widehat{m}c^{2}}{k_{B}T},
\label{eq-R-define}\end{equation}
where $\Gamma$ is the adiabatic index, $\widehat{m}$ is the average particle
mass and $k_{B}$ is the Boltzmann constant.  $\mathscr{R}^{-1}$ is therefore a
measure of the temperature. We make the approximation that the external medium
around the jet is isothermal, so $\mathscr{R}_{e}$ is constant. There is
evidence for a temperature gradient on the relevant scales \citep{hardcastle02},
but the isothermal approximation has a very small effect on our results since
the energy entrained from the external medium is negligible (B94, LB02b) and
$\mathscr{R}_{e} \gg 1$ (Section~\ref{energy}).

\subsection{Conservation laws for the flaring region}

The main difference between our work and that of B94 and LB02b is that we divide the
flaring region into two parts: the laminar jet and the shear layer. Thus our
conservation equations include distinct terms associated with each of these
components.

\subsubsection{Conservation of rest mass}

We use the following notations: $r_{j}$, $r_{s}$ are the radii of the laminar jet and the shear layer respectively, $\rho$ is the proper density, $v$ is the bulk velocity, $\beta = v/c$ and $\gamma=(1-\beta^2)^{-1/2}$ is the bulk Lorentz factor. The rest mass of the material passing through the total jet cross section $A(x)=\pi r_{s}(x)^{2}$ per unit time is equal to the rest mass of the material
entering through the cross section 0 plus the total entrained mass from the
environment. We express the mass fluxes \textbf{$\dot{M}$}, in the laminar jet and the shear layer at distance $x$ separately by:
{\setlength\arraycolsep{1pt}
\begin{eqnarray}
& &\dot{M}_{j}(x)=\gamma_{j}\rho_{j}(x)v_{j}\pi r_{j}(x)^{2}\\
& &\dot{M}_{s}(x)=\gamma_{s}\rho_{s}(x)v_{s}\pi \left [r_{s}(x)^{2}-r_{j}(x)^{2}\right ]
\end{eqnarray}
}
From equation (9) of B94,we have
\begin{equation}
\gamma_{j}\rho_{j,0}v_{j}\pi r_{0}^{2}+\int_{0}^{x}\rho_{e}(x')f(x')dx'
=\dot{M}_{j}(x)+\dot{M}_{s}(x) \label{eq-massconf}
\end{equation}
The first term on the left of equation~(\ref{eq-massconf}) is the rest mass of the material entering
through cross section 0 per unit time. The second term on the left is the
entrained mass flux. The terms on the right represent the rest masses of the
material passing through the cross sections of the laminar jet and the shear
layer per unit time at distance $x$.  We assume that the laminar jet
continuously supplies energy and momentum to the shear layer in such a way that
$\beta_{j}$ and $\beta_{s}$ remain constant throughout the flaring region. The
integral term $g_{\rm f}(x)=\int_{x_{0}}^{x}\rho_{e}(x')f(x')dx'$ is the mass
entrainment function, which was given in the form
$g_{\rm f}(x)=\int_{S}\rho\bm{v_{\rm ent}}\bm{\cdot n}dS$ by B94 ($\bm{n}$ is
the normal direction of the unit surface $dS$). $f(x)$ is therefore a function
that expresses the combination of the perpendicular entrainment velocity and the
shape of the jet boundary. The function $g_{\rm f}(x)$ is a measure of the total
mass entrained between the nucleus and distance $x$ per unit time.

We assume that the jet is in pressure equilibrium with the external medium
throughout the flaring and outer regions. Thus at fixed $x$, the pressures in the
laminar jet, the shear layer and the environment are all equal. Dividing by
$p(x)$ on both sides of equation~(\ref{eq-massconf}) and defining $F_{\rm f}(x)=cg_{\rm f}(x)/\left [\pi p(x)\right ]$, we get:
{\setlength\arraycolsep{1pt}
\begin{eqnarray}
&&\frac{\mathscr{R}_{j}\Gamma_{j}}{\Gamma_{j}-1}\gamma_{j}\beta_{j}\left [
    \frac{p_{0}}{p(x)}r_{0}^{2}-r_{j}(x)^{2}\right ]=\nonumber\\
&
&\frac{\mathscr{R}_{s}(x)\Gamma_{s}}{\Gamma_{s}-1}\gamma_{s}\beta_{s}\left [r_{s}(x)^{2}-r_{j}(x)^{2}\right ]-F_{\rm f}(x).
\label{eq-mass-flare}\end{eqnarray}
} 

\begin{table*}
 \centering
\caption{Definitions of key parameters and functions. Columns 4 and 6 indicate whether the
  values are assumed a priori, inferred from fits of relativistic flow models to
  radio images (`Radio'), derived from X-ray observations of the surrounding hot gas
  (`X-ray') or calculated.}
\begin{tabular}{cccccc}
\hline \hline
 & & \multicolumn{2}{c}{Flaring region} & \multicolumn{2}{c}{Outer region}\\
Name & Physical meaning & value & origin & value & origin\\
\hline
$\Gamma_{j}$ & adiabatic index of the laminar jet & 4/3, constant & assumed & - & -\\
$\Gamma_{s}$ & adiabatic index of the shear layer & 4/3, constant & assumed & 4/3, constant & assumed\\
$\Gamma_{e}$ & adiabatic index of the environment & 5/3, constant & assumed & 5/3, constant & assumed\\
$\beta_{j}$ & bulk velocity of the laminar jet & constant & Radio & - & -\\
$\beta_{s}$ & bulk velocity of the shear layer & constant & Radio & function of $x$ & calculated\\
$\beta_{1}$ & the bulk velocity at the beginning point of the outer region & - & - & constant & Radio\\
$\mathscr{R}_{j}$ & ratio of rest mass energy to non-relativistic enthalpy for laminar jet& constant & calculated & - & -\\
$\mathscr{R}_{s}$ & ratio of rest mass energy to non-relativistic enthalpy for shear layer& function of $x$ & calculated & function of $x$ & calculated\\
$\mathscr{R}_{1}$ & the value of $\mathscr{R}_{s}$ on the cross section 1& - & - & constant & calculated\\
$p$ & external pressure on cross section $x$ & function of $x$ & X-ray & function of $x$ & X-ray\\
$r_{j}$ & the radius of the laminar jet & function of $x$ & calculated & - & -\\
$r_{s}$ & the radius of the shear layer & function of $x$ & Radio & function of $x$ & Radio\\
$r_{0}$ & the jet radius at the brightening point & constant & Radio & - & -\\
$r_{1}$ & the shear layer radius at the beginning point of the outer region & - & - & constant & Radio \\
$g_{\rm f}$ & entrained mass per time from cross section 0 up to cross section $x$ & function of $x$ & calculated & - & -\\
$g_{\rm o}$ & entrained mass per time from cross section 1 up to cross section $x$ & - & - & function of $x$ & calculated\\
\hline
\end{tabular}
\label{parameter1}
\end{table*}

\subsubsection{Conservation of momentum}

The momentum flow through the cross section $A(x)$ per unit time should be equal
to the momentum of the material coming out of the initial cross section $0$ per
unit time, modified by the effects of buoyancy and differences in pressure
between the flow and its environment. We express the momentum flux, $\dot{P}$, by:
{\setlength\arraycolsep{1pt}
\begin{eqnarray}
& & \dot{P}_{j}(x)=\left [\gamma_{j}^{2}\frac{\omega_{j}(x)}{c^{2}}v_{j}^{2}+\Delta p_{j,l}(x)\right ]\pi r_{j}(x)^{2},\\ 
& & \dot{P}_{s}(x)=\left [\gamma_{s}^{2}\frac{\omega_{s}(x)}{c^{2}}v_{s}^{2}+\Delta p_{l,e}(x)\right ]\pi\left [r_{s}(x)^{2}-r_{j}(x)^{2}\right ],
\end{eqnarray}
}
where $\Delta p_{j,l}(x)=p_{j}(x)-p_{s}(x)$ and $\Delta
p_{l,e}(x)=p_{s}(x)-p_{e}(x)$ are the pressure differences at distance
$x$. In our pressure-matched case, they are all equal to 0. We assume that
material is entrained from the environment 
with a small bulk velocity and therefore that it contributes negligible momentum
compared with that of the jet. We rewrite equation (16) in B94 for our case as:
\begin{equation}
\gamma_{j}^{2}\frac{\omega_{j,0}}{c^{2}}v_{j}^{2}\pi r_{0}^{2} =
\dot{P}_{j}(x)+\dot{P}_{s}(x)+\phi(x), \label{eq-momconf}
\end{equation}
where
$\phi(x)=\int_{x_{0}}^{x}dx'\left [\frac{dp_{e}}{dx'}\int_{A}(1-\frac{\rho_{j}}{\rho_{e}})dS\right ]$
is the buoyancy term and for $\rho_{j}\ll\rho_{e}$ in our model, we have
$\phi(x)=\int_{x_{0}}^{x}\pi r_{s}(x')^{2} dp$. The momentum equation can then be simplified to:
{\setlength\arraycolsep{1pt}
\begin{eqnarray}
&
  &\frac{(\mathscr{R}_{j}+1)\Gamma_{j}}{\Gamma_{j}-1}\gamma_{j}^{2}\beta_{j}^{2}
  \left [\frac{p_{0}}{p(x)}r_{0}^{2}-r_{j}(x)^{2}\right ]=\nonumber\\
& &\frac{(\mathscr{R}_{s}(x)+1)\Gamma_{s}}{\Gamma_{s}-1}\gamma_{s}^{2}\beta_{s}^{2}\left [r_{s}(x)^{2}-r_{j}(x)^{2}\right ]+\frac{\phi(x)}{\pi p(x)}.
\label{eq-mom-flare}\end{eqnarray}
}
\subsubsection{Conservation of energy}
\label{energy}

The energy passing through the jet cross section must also be conserved. We
express the energy flux (or jet power), $Q$, at distance $x$ for the two regions as:
{\setlength\arraycolsep{1pt}
\begin{eqnarray}
& &Q_{j}(x)=\gamma_{j}^{2}\omega_{j}(x)v_{j}\pi r_{j}(x)^2, \label{eq-jet-energy-flux}\\
& &Q_{s}(x)=\gamma_{s}^{2}\omega_{s}(x)v_{s}\pi \left [r_{s}(x)^{2}-r_{j}(x)^{2}\right ].
\end{eqnarray}
}
B94 gives the relevant conservation law in his equation (26) and we rewrite this as:
\begin{equation}
\gamma_{j}^{2}\omega_{j,0}v_{j}\pi r_{0}^{2}+\int_{x_{0}}^{x}\omega_{e}(x')f(x')dx'=Q_{j}(x)+Q_{s}(x).\label{eq-enconf}
\end{equation}
As the environment is dominated by the rest mass energy, so
$\mathscr{R}_{e}$ is extremely large, and we can approximate $1+1/\mathscr{R}_{e}\approx1$ at all position. Thus, $\int_{x_{0}}^{x}\omega_{e}(x')f(x')dx'=\int_{x_{0}}^{x}c^{2}\left [1+1/\mathscr{R}_{e}(x')\right ]f(x')dx'\approx c^{2}g_{\rm f}(x)$.
Dividing both sides by $cp(x)$, we get:
{\setlength\arraycolsep{1pt}
\begin{eqnarray}
&
  &\frac{(\mathscr{R}_{j}+1)\Gamma_{j}}{\Gamma_{j}-1}\gamma_{j}^{2}\beta_{j}
  \left [\frac{p_{0}}{p(x)}r_{0}^{2}-r_{j}(x)^{2}\right ]=\nonumber\\
& &\frac{\left [\mathscr{R}_{s}(x)+1\right ]\Gamma_{s}}{\Gamma_{s}-1}\gamma_{s}^{2}\beta_{s}\left [r_{s}(x)^{2}-r_{j}(x)^{2}\right ]-F_{\rm f}(x).
\label{eq-en-flare}\end{eqnarray}
}
\subsection{Conservation laws for the outer region}
\label{cons-outer}

For the outer region the conservation equations are similar, but without the
laminar jet term. Another important difference is that the initial cross section
is now at the end of the flaring region (point 1 in Fig.~\ref{cartoon}). The
entrained mass and energy now denote the values integrated from point 1
(distance $x_{1}$) up to distance $x$. Finally, the velocity of the layer,
$\beta_{s}$, is a function of distance $x$. The three equations analogous to
equations~(\ref{eq-massconf}), (\ref{eq-momconf}), and (\ref{eq-enconf}) are
then given by

{\setlength\arraycolsep{1pt}
\begin{eqnarray}
&&\gamma_{1}\rho_{1}v_{1}\pi r_{1}^{2}=\dot{M}_{s}(x)-g_{\rm o}(x), \label{eq-massconf-out}\\
&&\gamma_{1}^{2}\frac{\omega_{1}}{c^{2}}v_{1}^{2}\pi
  r_{1}^{2}=\dot{P}_{s}(x)+\phi(x), \label{eq-momconf-out}\\
&&\gamma_{1}^{2}\omega_{1}v_{1}\pi r_{1}^{2}=Q_{s}(x)-\int_{x_{1}}^{x}\omega_{e}(x')f(x')dx'.\label{eq-enconf-out}
\end{eqnarray}
}
The term $g_{\rm o}(x)=\int_{x_{1}}^{x}\rho_{e}(x')f(x')dx'$ is equal
to the amount of entrained mass per unit time. With the same
definitions of $F(x)$ and $R$ as given above, these three equations
can be written in the following ways

{\setlength\arraycolsep{1pt}
\begin{eqnarray}
\frac{\Gamma_{s}\gamma_{1}\beta_{1}}{\Gamma_{s}-1}\frac{p_{1}r_{1}^{2}}{p(x)}&=&\frac{\Gamma_{s}}{\Gamma_{s}-1}
\frac{\mathscr{R}_{s}(x)}{\mathscr{R}_{1}}\gamma_{s}(x)\beta_{s}(x)r_{s}(x)^{2}-\frac{F_{\rm o}(x)}{\mathscr{R}_{1}},
\label{eq-mass-out}\\
\frac{\Gamma_{s}\gamma_{1}^{2}\beta_{1}^{2}}{\Gamma_{s}-1}\frac{p_{1}r_{1}^{2}}{p(x)}&=&\frac{\Gamma_{s}}{\Gamma_{s}-1}
\frac{\mathscr{R}_{s}(x)+1}{\mathscr{R}_{1}+1}\gamma_{s}(x)^{2}\beta_{s}(x)^{2}r_{s}(x)^{2}+\nonumber\\
&&\frac{1}{\mathscr{R}_{1}+1}\frac{\phi(x)}{\pi p(x)},\label{eq-mom-out}\\
\frac{\Gamma_{s}\gamma_{1}^{2}\beta_{1}}{\Gamma_{s}-1}\frac{p_{1}r_{1}^{2}}{p(x)}&=&
\frac{\Gamma_{s}}{\Gamma_{s}-1}\frac{\mathscr{R}_{s}(x)+1}{\mathscr{R}_{1}+1}\gamma_{s}(x)^2\beta_{s}(x)r_{s}(x)^{2}-\frac{F_{\rm o}(x)}{\mathscr{R}_{1}+1}.\nonumber\\
\label{eq-en-out}
\end{eqnarray}
}

\begin{figure}
\includegraphics[width=0.5\textwidth]{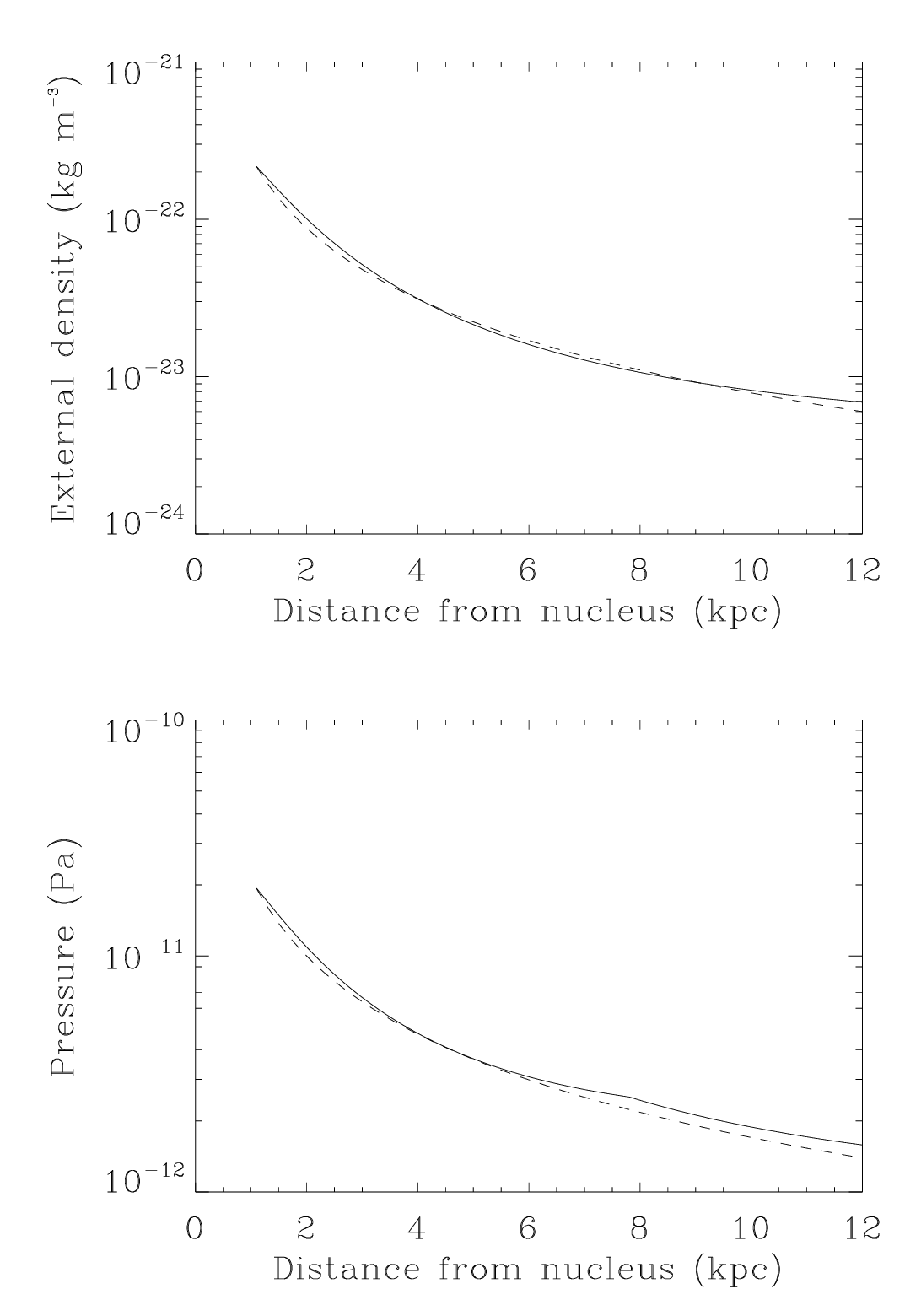}
\caption{The external density and pressure profiles for 3C\,31. The solid lines
are derived from the double-beta-model fit to the number density and pressure [equations~(\ref{eq-doublebeta-density}) and (\ref{eq-doublebeta-pressure})]
while the dashed lines are power-law approximations with indices of
$\alpha_{1}=1.5$ and $\alpha_{2}=1.1$, as described in the text.}
\label{density}
\end{figure}

\section{Solutions}
\label{solutions}

In this section, we will solve equations~(\ref{eq-mass-flare}),
(\ref{eq-mom-flare}), (\ref{eq-en-flare}) for the flaring region, and
equations~(\ref{eq-mass-out}), (\ref{eq-mom-out}), (\ref{eq-en-out}) for the
outer region in terms of quantities which can be inferred either from fits of
relativistic flow models to radio images (jet and layer velocities in the
flaring region, together with the radius of the layer in both regions) or from
X-ray observations of the surrounding hot gas (external density, temperature and
pressure). We can then derive the shape of the laminar jet, $r_{j}(x)$, the
variation of velocity with distance in the outer region, $\beta_{s}(x)$, the
values of $\mathscr{R}$ in the various regions, the entrainment function and the
velocity of entrainment.

We assume that the laminar jet has a relativistic equation of state with
$\Gamma_{j}=4/3$; the environment has $\Gamma_{e}=5/3$. The shear layer contains
mixed material but the energy density must still be dominated by relativistic
particles (B94), and we therefore take $\Gamma_{s}=4/3$.

\subsection{Solutions for the flaring region}

From equations~(\ref{eq-mass-flare}) and (\ref{eq-en-flare}), we have:
\begin{equation}
\frac{r_{s}(x)^{2}-r_{j}(x)^{2}}{\frac{p_{0}}{p(x)}r_{0}^{2}-r_{j}(x)^{2}}=\frac{\frac{\Gamma_{j}}{\Gamma_{j}-1}\gamma_{j}\beta_{j}\left [\mathscr{R}_{j}-(\mathscr{R}_{j}+1)\gamma_{j}\right ]}{\frac{\Gamma_{s}}{\Gamma_{s}-1}\gamma_{s}\beta_{s}\{\mathscr{R}_{s}(x)-\left [\mathscr{R}_{s}(x)+1\right ]\gamma_{s}\}}.
\end{equation}
At the same time, from equation~(\ref{eq-mom-flare}), we have:
{\setlength\arraycolsep{0pt}
\begin{eqnarray}
&&\frac{r_{s}(x)^{2}-r_{j}(x)^{2}}{\frac{p_{0}}{p(x)}r_{0}^{2}-r_{j}(x)^{2}}=\frac{\frac{\Gamma_{j}}{\Gamma_{j}-1}(\mathscr{R}_{j}+1)\gamma_{j}^2\beta_{j}^2-\frac{\phi(x)}{p(x)r_{s}(x)^2-p_{0}r_{0}^2}}{\frac{\Gamma_{s}}{\Gamma_{s}-1} \left [(\mathscr{R}_{s}(x)+1\right ]\gamma_{s}^2\beta_{s}^2-\frac{\phi(x)}{p(x)r_{s}(x)^2-p_{0}r_{0}^2}}.\nonumber\\
&&\label{eq-flare}
\end{eqnarray}
} Thus, we can express $\mathscr{R}_{s}$ as a function of
$\mathscr{R}_{j}$, $\beta_{j}$, $\beta_{s}$ and the buoyancy
term, $\phi(x)$, which can be calculated from the pressure profile and the
shape of the jet $r_{s}(x)$:
\begin{equation}
\mathscr{R}_{s}(x)=\frac{C(x)+B(x)\gamma_{s}}{D(x)-A\gamma_{s}\beta_{s}},
\label{eq-r-flare}\end{equation}
where
{\setlength\arraycolsep{1pt}
\begin{eqnarray}
&&A = \mathscr{R}_{j}-(\mathscr{R}_{j}+1)\gamma_{j},\\
&&B(x) = (R_j+1)\gamma_{j}\beta_{j}+\frac{\Gamma_{j}-1}{\Gamma_{j}\gamma_{j}\beta_{j}}\frac{\phi(x)}{p(x)\pi r_{s}(x)^{2}-p_{0}\pi r_{0}^{2}}\\
&&C(x) = A\left [\gamma_{s}\beta_{s}+\frac{\Gamma_{s}-1}{\Gamma_{s}\gamma_{s}\beta_{s}}\frac{\phi(x)}{p(x)\pi r_{s}(x)^{2}-p_{0}\pi r_{0}^{2}}\right ]\\
&&D(x) = B(x)(1-\gamma_{s}).
\end{eqnarray}
} Also, from equations~(\ref{eq-mass-flare}) and (\ref{eq-flare}), we can express the shape of the
laminar jet and the entrainment function by:
{\setlength\arraycolsep{1pt}
\begin{eqnarray}
r_{j}(x)^{2} & = & \frac{p_{0}r_{0}^{2}}{p(x)}-\frac{ r_{s}(x)^{2}\{\dot{P}_{s}(x)\left [\frac{p_{0}r_{0}^{2}}{p(x)r_{s}(x)^{2}}-1\right ]+\tau(x)\phi(x)\}}{{\dot{P}_{s}(x)}-\dot{P}_{j}(x)\frac{\tau(x)}{\kappa(x)}},\nonumber\\
\label{eq-laminar-shape}\\
g_{\rm f}(x) & = &  \frac{\left [1-\frac{p_{0}r_{0}^{2}}{p(x)r_{s}(x)^{2}}\right ](\beta_{j}-\beta_{s})+c\phi(x)\left [\frac{\kappa(x)}{Q_{j}(x)}-\frac{\tau(x)}{Q_{s}(x)}\right ]}{c^{2}\left [\frac{\beta_{j}\tau(x)}{Q_{s}(x)}-\frac{\beta_{s}\kappa(x)}{Q_{j}(x)}\right ]},\nonumber\\
\end{eqnarray}
} 
where $\kappa(x)=\left [r_{j}(x)/r_{s}(x)\right ]^{2}$ and $\tau(x)=1-\kappa(x)$ are the
fractions of jet and shear layer, respectively, at distance $x$. Although the
expressions for $\dot{P}_{s}(x)$ and $\dot{P}_{j}(x)$ contain $r_{j}(x)$
[equation~(\ref{eq-laminar-shape})], $\dot{P}_{s}(x)/\tau(x)$ and
$\dot{P}_{j}(x)/\kappa(x)$ are functions only of observable parameters together
with $\mathscr{R}_{s}(x)$ and $\mathscr{R}_{j}$. By applying the boundary
condition $r_{j}(x_{1})=0$, we can derive $\mathscr{R}_{j}$ and then solve for
$\mathscr{R}_{s}(x)$ from equation~(\ref{eq-r-flare}) given the shape of the
outer boundary of the shear layer, $r_{s}(x)$. Finally, we can determine the
shape of the laminar jet boundary, and the function $F(x)$, which can then be
used to calculate the entrainment function.

\begin{figure*}
\includegraphics[width=\textwidth]{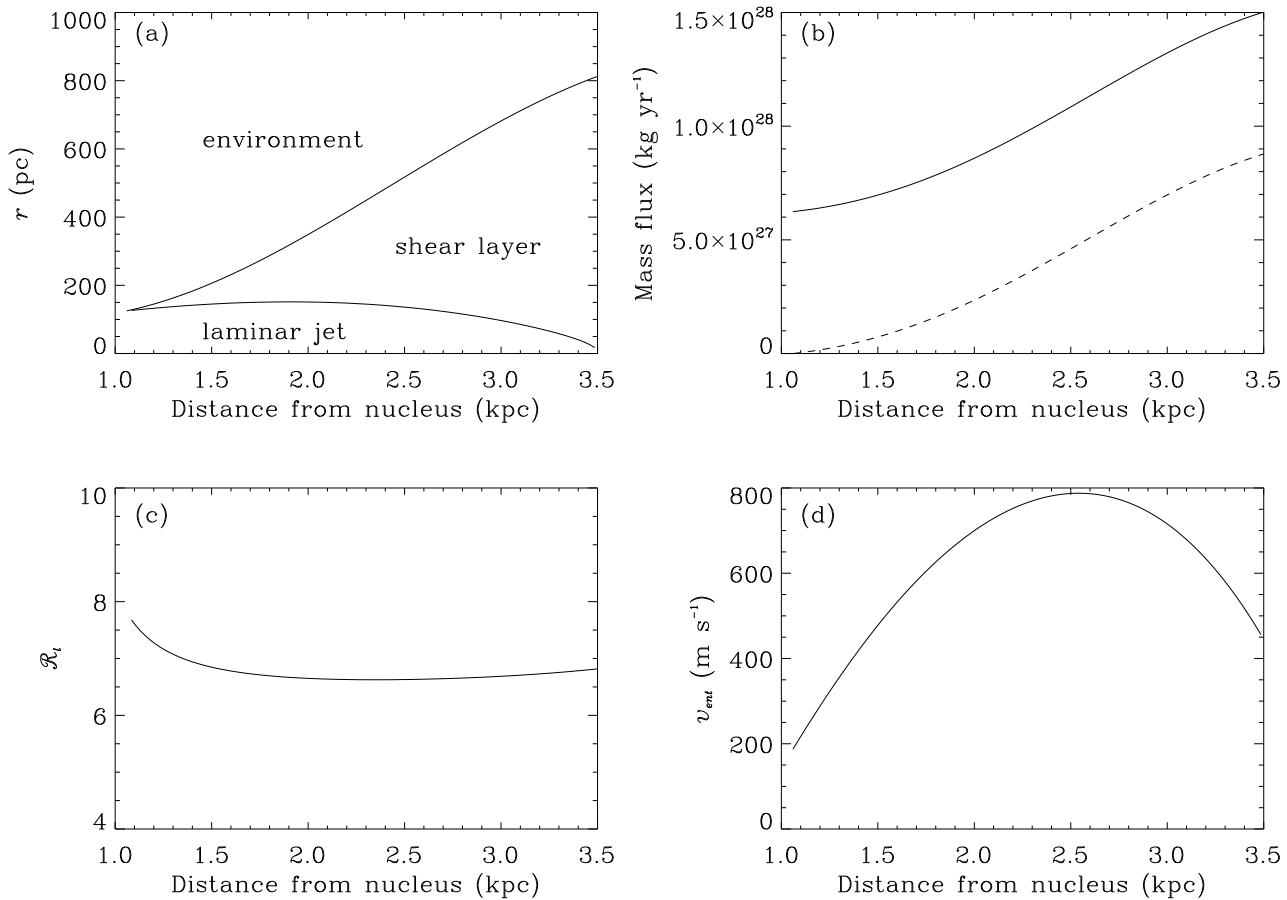}
\caption{Results from our model for the flaring region of 3C\,31. (a)
  Geometry. The outer edge of the flow and the boundary between the laminar core
  and shear layer are shown. (b) Mass flux at distance $x$. The full and dashed lines indicate
the total mass flux and the contribution from entrainment, respectively. (c)
  Profile of $\mathscr{R}_{s}(x)$. (d) The entrainment velocity perpendicular to the
outer boundary at distance $x$.} \label{flare}
\end{figure*}

\subsection{Solutions for the outer region}

In the outer region there is no laminar jet to supply energy to the shear layer
but matter continues to be entrained from the environment. Thus both
$\beta_{s}$ and $\mathscr{R}_{s}$ are expected to be functions of $x$. We solve
the equations numerically, using the following steps. Equations~(\ref{eq-mass-out}) and (\ref{eq-en-out})
give:
\begin{equation}
\mathscr{R}_{s}(x)=\frac{\frac{\dot{M}_{1}\mathscr{R}_{s}(x)}{\dot{M}_{s}(x)\mathscr{R}_{1}}\left [\mathscr{R}_{1}(\gamma_{1}-1)+\gamma_{1}\right ]-\gamma_{s}(x)}{\gamma_{s}(x)-1},
\label{eq-r-out1}\end{equation}
while equation~(\ref{eq-mom-out}) gives:
\begin{equation}
\mathscr{R}_{s}(x)=\frac{\Gamma_{s}-1}{\Gamma_{s}}\frac{\dot{P}_{1}-\pi p(x)\phi(x)}{\dot{P}_{s}(x)/\left [\mathscr{R}_{s}(x)+1\right ]}-1.
\label{eq-r-out2}\end{equation}
Again, $\mathscr{R}_{s}(x)$ occurs on the right-hand sides of
equation~(\ref{eq-r-out1}) and (\ref{eq-r-out2}), but
$\dot{M}_{s}(x)/\mathscr{R}_{s}(x)$ and $\dot{P}_{s}(x)/[\mathscr{R}_{s}(x)+1]$ are just functions of $\beta_{s}(x)$ and
other observable parameters. Combining these two equations, we can solve
numerically for the value of $\beta_{s}(x)$: the shape of the boundary,
$r_{s}(x)$, is constrained from observations, so the only unknown parameters are
$\beta_{s}(x)$, which in turn determines $\gamma_{s}(x)$. Then, with the known
value of $\beta_{s}(x)$, we can express the entrainment function as follows.
\begin{eqnarray}
g_{\rm o}(x)=\frac{\dot{M}_{1}}{\mathscr{R}_{1}}\frac{\gamma_{1}(\mathscr{R}_{1}+1)-\left [\frac{\dot{M}_{1}\mathscr{R}_{s}(x)}{\dot{M}_{s}(x)\mathscr{R}_{1}}+1\right ]\gamma_{s}(x)}{\gamma_{s}(x)-1}.
\end{eqnarray}
Observations show that the radius of the shear layer in the outer region
$r_{s}(x)$ increases linearly with $x$. We use this observed variation as the
input function and predict the distributions of $\beta_{s}(x)$,
$\mathscr{R}_{s}(x)$ and $g_{\rm o}(x)$

\begin{figure*}
\includegraphics[width=\textwidth]{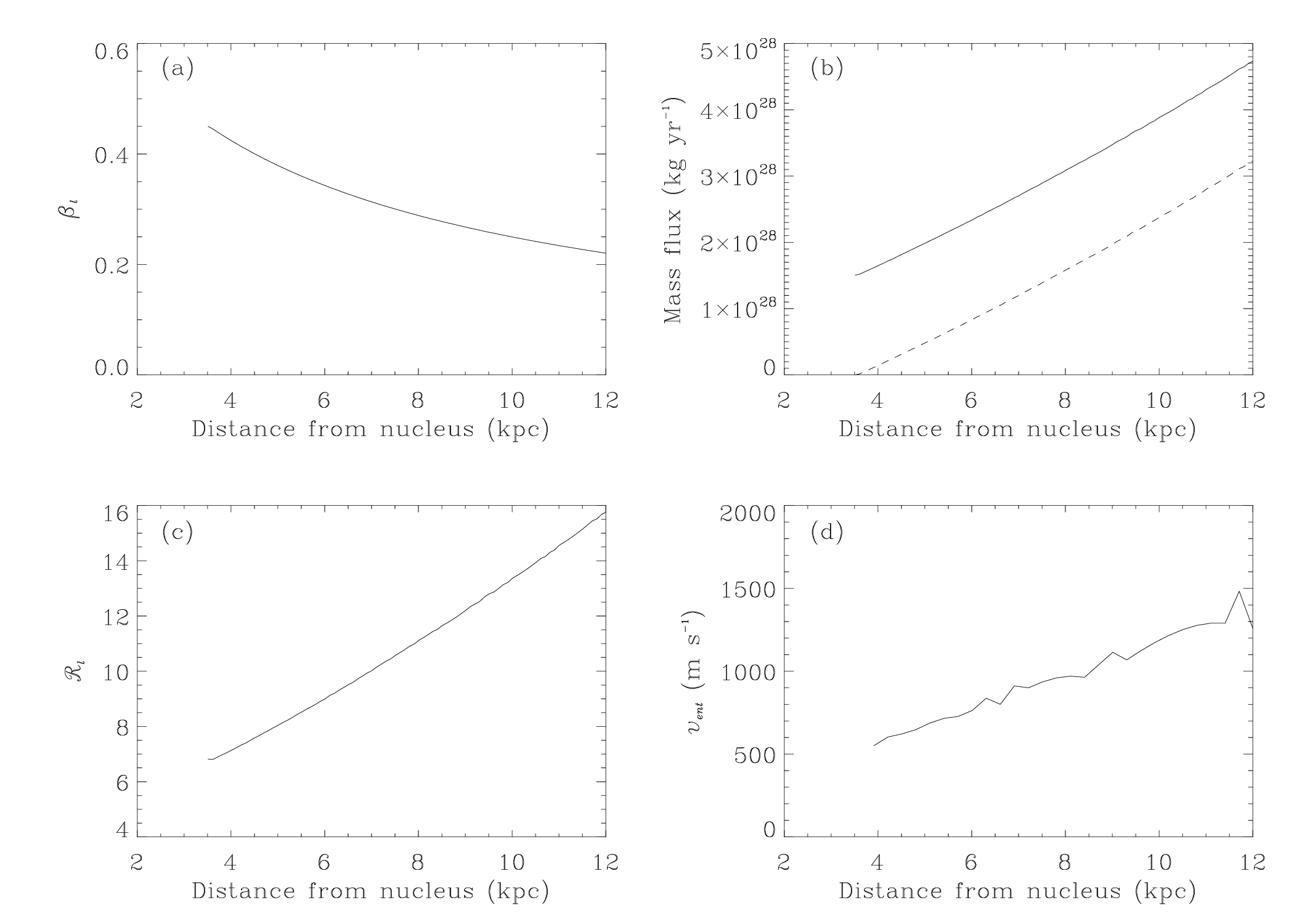}
\caption{Results from our model for the outer region of 3C\,31.
(a) Profile of bulk velocity $\beta_{s}(x)$. (b) Mass flux 
$\dot{M}(x)$ at distance $x$. The full and dashed lines indicate the total mass
  flux and the contribution from 
entrainment, respectively.  (c) Profile of $\mathscr{R}_{s}(x)$. (d) 
The entrainment velocity perpendicular to the outer boundary as a function of 
distance, $x$. The jagged shape of the profile is a numerical artefact, but the
overall shape is correct.} \label{outer}
\end{figure*}


\subsection{Summary of the solutions}
\label{parameter}

To get solutions for both the flaring region and the outer region, we adopt the
shape function $r_{s}(x)$ from model-fitting to radio images, together with the
velocities $\beta_{s}$ and $\beta_j$ for the flaring region. We also adopt the pressure profiles from X-ray observations. This leaves three functions which
need to be evaluated at each distance $x$: $r_{j}(x)$, $F(x)$ and
$\mathscr{R}_{s}(x)$ for the flaring region, and $\beta_{s}(x)$,
$\mathscr{R}_{s}(x)$ and $F(x)$ for the outer region.

The three equations from the conservation laws thus form a closed system. The
input and derived parameters are listed in Table \ref{parameter1}.

\section{Application to 3C\,31}
\label{3c31}

Having established a system of equations which describe the structure and
kinematics of an FR\,I jet, we now compare the results with observational data
and models for the well-observed source 3C\,31. Geometrical (projection factor
and radius) and velocity information are inferred from the relativistic-flow
models of LB02a. Fits to the density, temperature and pressure of the hot gas
surrounding the jets are as given by \citet{hardcastle02} and used in the
quasi-one-dimensional conservation-law analysis of LB02b. As in these
references, we adopt a concordance cosmology with Hubble constant, $H_0$ =
70\,$\rm{km\,s^{-1}\,Mpc^{-1}}$, $\Omega_\Lambda = 0.7$ and $\Omega_M =
0.3$. At the redshift of the host galaxy of 3C\,31, $z = 0.0169$, this gives a
scale of 0.344\,kpc\,arcsec$^{-1}$.

\subsection{Inferences from observation}
\label{input-parms}

The parameters defining the edge of the shear layer projected on the sky are
determined by fitting to the total-intensity distribution. The angle to the line
of sight required to correct for projection (52$^\circ$ for 3C\,31) is derived
from the relativistic-flow model.  In LB02a, the shape of the shear layer in the
flaring region is described by the polynomial $r_{s}(x)=a+bx+cx^2+dx^3$ with
$r_{0}=0.125$\,kpc at 1.1\,kpc and $r_{1}=0.815$\,kpc at 3.5\,kpc. The shear
layer initially expands slowly, then goes through a phase of faster expansion
before recollimating at the end of the flaring region. In the outer region, the
shear layer expands conically with an intrinsic half-angle of $13.1^{\circ}$.
At the beginning of the flaring region, we assume that there is no shear layer,
so we use the on-axis bulk velocity inferred by LB02a to characterize the jet,
$v_{j}=0.77c$. We suppose that the shear layer makes up essentially all of the
flow at the end of the flaring region. LB02a infer a variation of velocity
across the flow from $0.37c$ -- $0.55c$ at this distance so we take a
representative value of $v_{s}=0.45c$.

\citet{hardcastle02} have estimated  the external density and pressure
profiles for 3C\,31 from X-ray observations. The density profile is given by:
\begin{equation}
\rho_{e}(x)=m_{p}n_{\rm{e}}(x)/\chi_{H},
\end{equation}
where $m_{p}$ is the mass of a proton, $\chi_{H}=0.74$ is the abundance
of hydrogen by mass and $n_{e}(x)$ is the proton number density of the
environment given by:
\begin{equation}
n_{e}(x)=n_{c}(1+x^{2}/x_{c}^{2})^{-3\beta_{c}/2}+n_{g}(1+x^{2}/x_{g}^{2})^{-3\beta_{g}/2}.
\label{eq-doublebeta-density}\end{equation}
The numerical values of the parameters are: $n_{c}=1.8\times10^{5}$\,m$^{-3}$,
$n_{g}=1.9\times10^{3}$\,m$^{-3}$, $\beta_{c}=0.73$, $\beta_{g}=0.38$,
$x_{c}=1.2$\,kpc, $x_{g}=52$\,kpc. The temperatures estimated by
\citet{hardcastle02} range from $4.9 \times 10^6$\,K to $1.7 \times 10^7$\,K,
corresponding to $\mathscr{R}_{e} = 5 \times 10^5$ to $1.5 \times 10^5$. Thus
the approximation $1+1/\mathscr{R}_{e}\approx1$ (Section 3.1.3) is valid to high accuracy.
The pressure is given by \citet{BW93}:
\begin{equation}
p(x)=kT(x)n_{e}(x)/(\mu\chi_{H}),\label{eq-doublebeta-pressure}
\end{equation}
where $\mu=0.6$ is the mass per particle. For simplicity, we approximate the pressure and density distributions using power-law forms:
{\setlength\arraycolsep{1pt}
\begin{eqnarray}
&&\rho_{e}(x)=\rho_{e,0}(\frac{x}{x_{0}})^{-\alpha_1},\\
&&p(x)=p_{0}(\frac{x}{x_{0}})^{-\alpha_2},
\end{eqnarray}
} where $x_{0}$ is the position of the brightening point.
$\rho_{e,0}=2.16\times10^{-22}$\,kg\,m$^{-3}$ and $p_{0}=1.93\times10^{-11}$\,Pa
are the density and pressure at $x_{0}$, respectively. The values
$\alpha_{1}=1.5$ and $\alpha_{2}=1.1$ give good approximations to the profiles,
and we adopt them in the following calculation. The corresponding density and
the pressure profiles are compared with those from \citet{hardcastle02} in
Figure \ref{density}.  Although we use an isothermal approximation in the
development of our model (Section~\ref{laws}), the assumed pressure profile includes
the effects of the temperature gradient.

\subsection{Results from the model}

\subsubsection{Flaring region}

With the parameters given in Section~\ref{input-parms}, we obtain
$\mathscr{R}_{j}=13.4$ in the flaring region. The profiles of
$\mathscr{R}_{s}(x)$ and the total mass flux passing through a given cross
section, $\dot{M}$, are plotted in Figure \ref{flare}. In the same figure, we
also plot $v_{\rm ent}$, the normal component of the entrainment velocity across
the surface of the jet. This is related to the entrainment function by
$v_{\rm ent}=(1/\rho_{e})dg/ds$.

The model predicts that the laminar jet initially expands at the beginning of
the flaring region and then starts to collapse $\approx$1.7\,kpc away from the
brightening point. Meanwhile, the value of $\mathscr{R}_{s}(x)$ drops a little
at the beginning of the flaring region and then reaches an asymptotic value of
$\approx$6.7. The initial decrease of $\mathscr{R}_{s}(x)$ occurs because the
small amount of entrained material at the beginning of the flaring region can
easily be heated by the the laminar jet.  The functional forms of
$\mathscr{R}_{s}(x)$ and $v_{\rm ent}(x)$ are constrained by the parameters
inferred for 3C\,31 and may differ in other sources. For example, if the shear
layer initially expands faster, $R_{s}(x)$ will be higher and $v_{\rm ent}$
lower throughout the flaring region.

\subsubsection{Outer region}

In the outer region, our model predicts that the bulk velocity $\beta_{s}$
should decrease smoothly with $x$. $\beta_{s} = 0.45$ at 3.5\,kpc, where it is
normalized to the mean value of the distribution derived by LB02a, decreasing to
$0.22$ at 12\,kpc. This is reasonably consistent with the velocity range derived
by LB02a ($\beta$ = 0.15 -- 0.22 at the same distance).  The value of $\mathscr{R}_{s}$
increases with $x$ in our solution, reflecting the increasing dominance of the
mass by entrained material. We plot $\mathscr{R}_{s}(x)$ and $\beta_{s}(x)$
together with profiles of mass flux and velocity in Figure \ref{outer}.

\subsubsection{Estimate of jet power}
\label{jetpower}

Using the calculated and observed parameters given above, we can estimate the
power of the jets in 3C\,31. The relevant parameter for comparison with
estimates by other methods (e.g.\ \citealt{birzan08}) is $\Phi$ (LB02b), the
energy flux of the jet with the rest-mass contribution subtracted. $\Phi = Q -
\dot{M}c^2$ in the notation of the present paper.  Applying
equation~(\ref{eq-jet-energy-flux}) at the brightening point, we get values of
$Q=3.4\times10^{37}$\,W and $\Phi = 1.6 \times 10^{37}$\,W.  3C\,31 is a fairly
powerful FR\,I source, with a monochromatic luminosity of 10$^{24.5}$\,W at
1.4\,GHz, approximately a factor of 10 below the FR\,I/FR\,II dividing line
plotted by \citet{lo96} given the absolute magnitude of the host galaxy
\citep{ol89}.  A total power of $\Phi=1.6\times10^{37}$\,W for the twin jets of
3C\,31 is well within the range derived from observations of cavities in the
X-ray gas surrounding other radio galaxies of comparable monochromatic
luminosity \citep{birzan08}.

\begin{figure}
\includegraphics[width=0.5\textwidth]{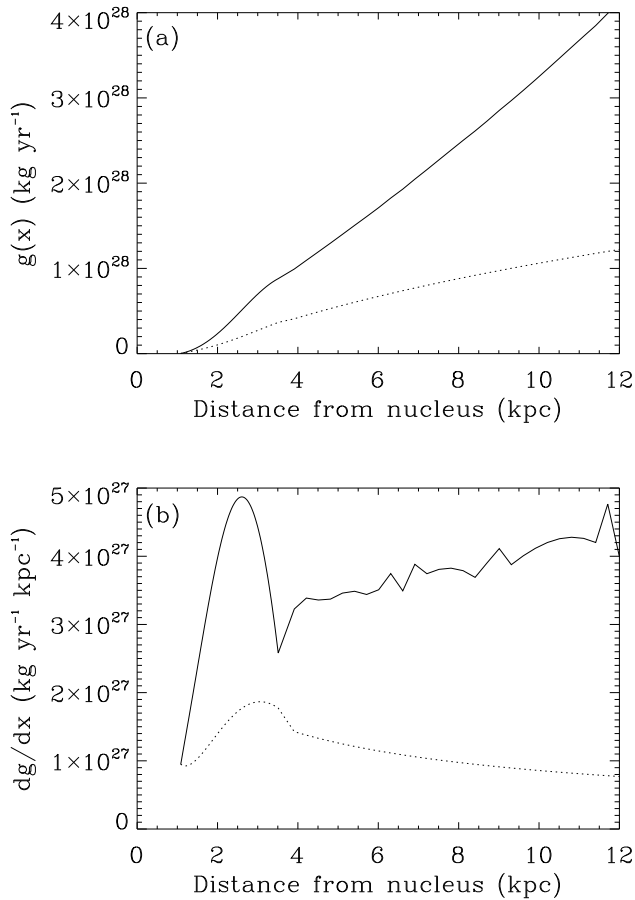}
\caption{(a) The entrainment function, $g(x)$, from our model (full line)
compared with the estimate from stellar mass loss within the jet, $g_{\rm s}(x)$
(dotted). (b) As in panel (a), but for the entrainment per unit length of the
jet, $dg/dx$.} \label{entrainment}
\end{figure}

\begin{figure}
\includegraphics[width=0.5\textwidth]{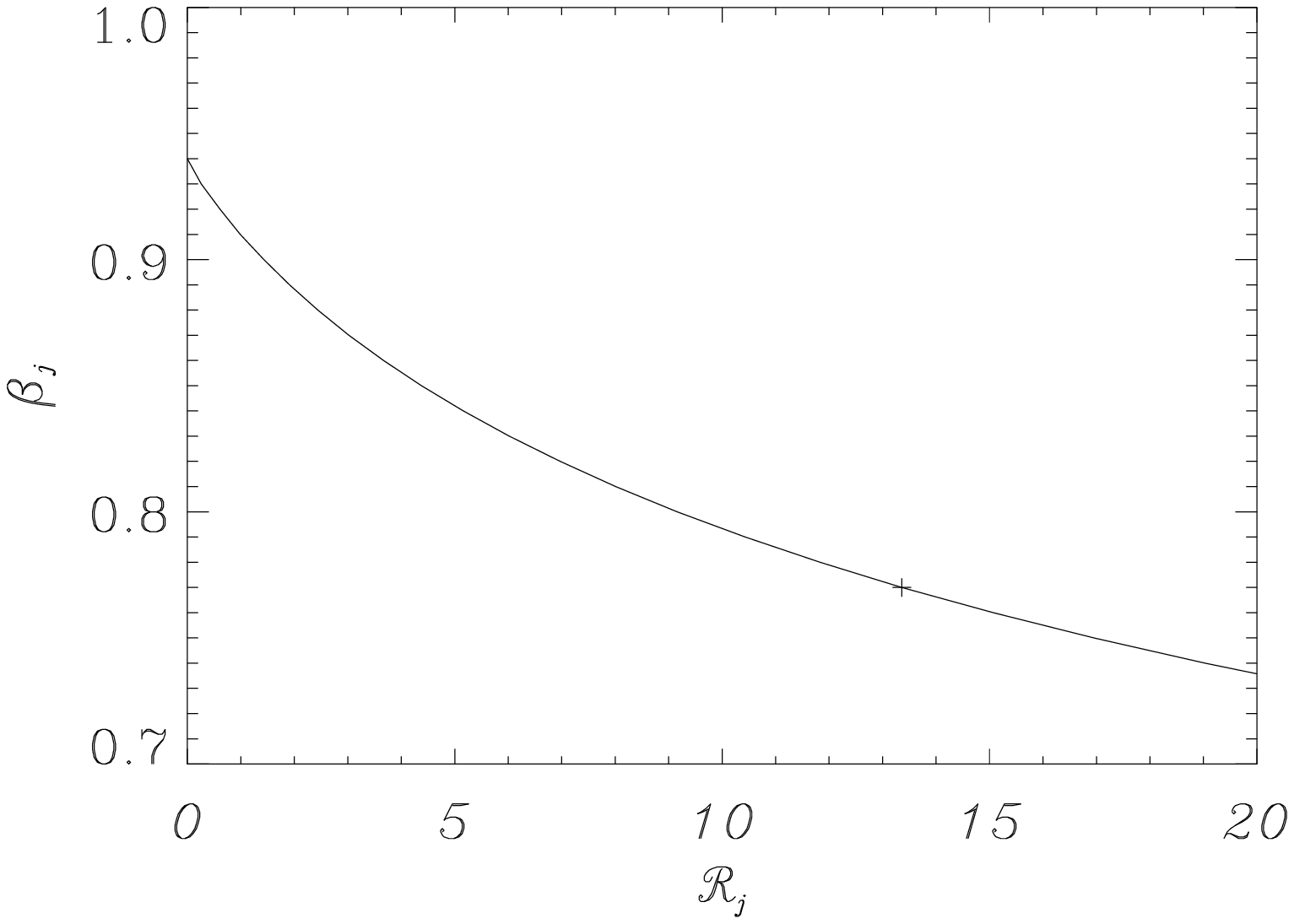}
\caption{The relation between $\mathscr{R}_{j}$ and $\beta_{j}$ for the flaring region. 
The values of $r_{0}$, $r_{1}$, $p(x)$ and $\beta_{s}$ are fixed at the values
determined for 3C\,31. The plus sign indicates the value of $\mathscr{R}_{j}$ for
3C\,31.} \label{rj_betaj}
\end{figure}

\begin{figure}
\includegraphics[width=0.5\textwidth]{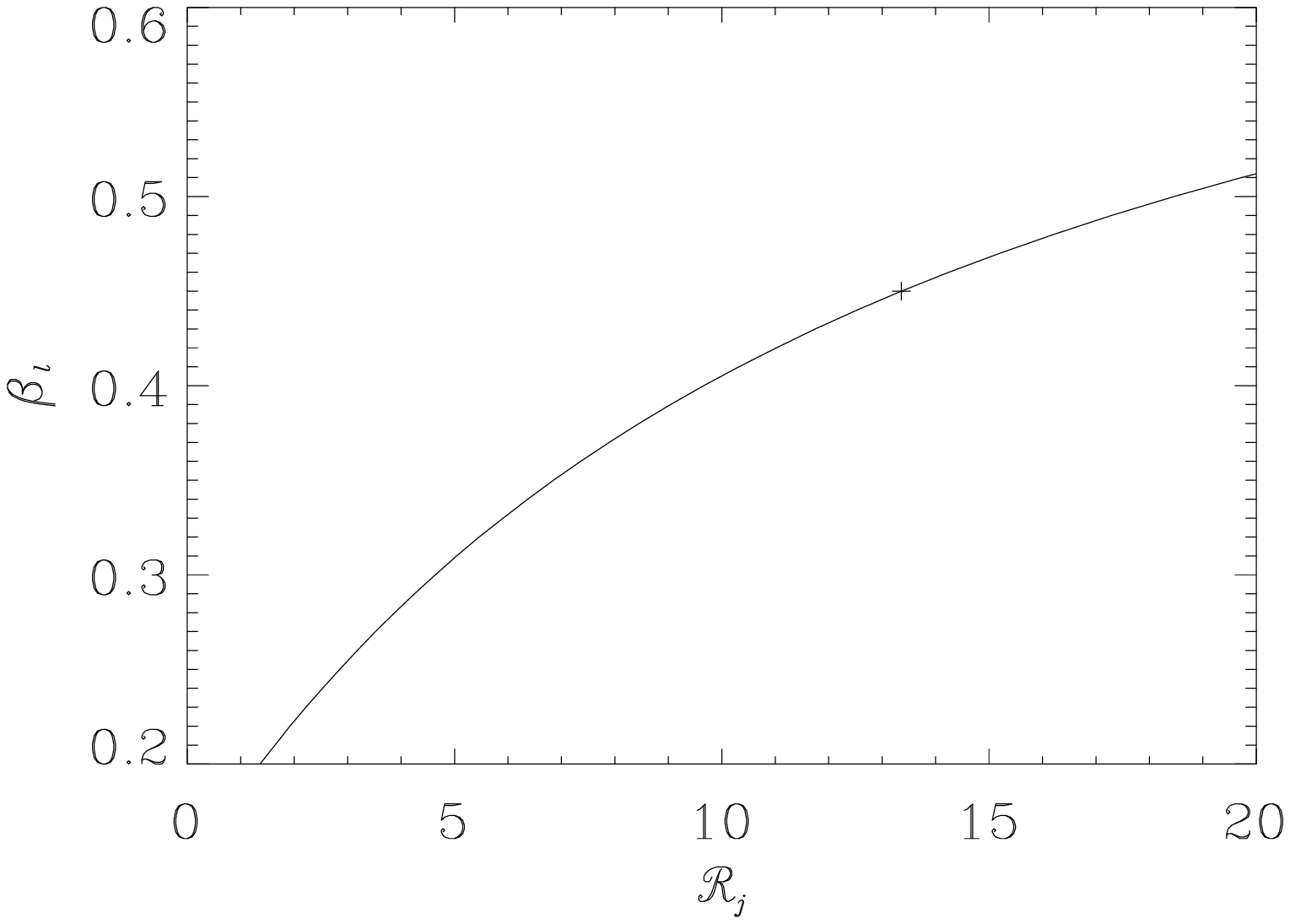}
\caption{The relation between $\mathscr{R}_{j}$ and $\beta_{s}$ for the flaring region. The values of
$r_{0}$, $r_{1}$, $p(x)$ and $\beta_{j}$ are set to the values determined for
  3C\,31. The plus sign indicates the value of $\mathscr{R}_{j}$ for 3C\,31} \label{rj_betal}
\end{figure}


\begin{figure}
\includegraphics[width=0.5\textwidth]{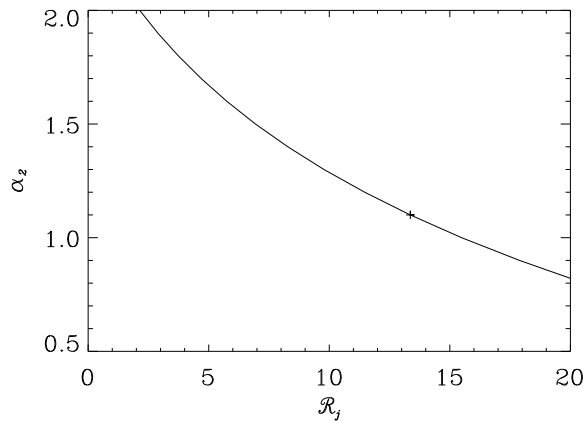}
\caption{The relation between $\mathscr{R}_{j}$ and $\alpha_2$ for the flaring region. The 
values of $r_{0}$, $r_{1}$, $\beta_{j}$ and $\beta_{s}$ are fixed at the values
determined for 3C\,31. The plus sign indicates the value of $\mathscr{R}_{j}$ for
3C\,31.} \label{rj_alpha}
\end{figure}

\subsubsection{Mass input from stellar mass loss}
\label{stars}

It has been argued that the deceleration in the flaring region could be caused
by the entrainment of stellar wind material from stars located inside the jet
\citep{komi94}. In order to test this idea, we adopt the estimate of mass input
from LB02b, who used a deprojection of R-band surface photometry for 3C\,31
\citep{ol89}, together with the same assumptions on conversion between stellar
luminosity and mass loss as in \citet{komi94} and \citet{bowman96}. The
corresponding entrainment per unit length (the derivative of the entrainment
function defined above) can be written as
\begin{equation}
dg_{\rm s}/dx =2.4\times10^{28}\pi
r_{s}(x)^{2}x^{-2.65}\textrm{\,kg\,kpc}^{-1}\textrm{\,yr}^{-1}, 
\end{equation}
where $r_{s}(x)$ and $x$ are in units of kpc. In Fig.~\ref{entrainment}, we
compare the entrainment function from our model and its derivative with 
those estimated for stellar mass loss.  At the beginning of the flaring region,
the stellar mass input rate is remarkably close to that required, given the
crudity of the assumptions. At larger distances, however, it falls well below
the level required to decelerate the jet. In the outer region, the entrainment
rate per unit length required by our model continues to increase, whereas that
from stellar mass loss decreases.  Thus, although stellar mass loss may be
important in initiating the jet deceleration at the start of the flaring region,
boundary-layer entrainment, as described by our model, is clearly required on
larger scales. Mass input distributed throughout the jet volume, as would be
expected from stellar mass loss, is a potential complication to our analysis.

\subsection{Comparison with LB02b}
\label{LB02compare}

It is of interest to compare the results of the present model with the
conservation-law analysis of LB02b. The treatments are very similar in many
respects, both relying on quasi-one-dimensional approximations and using
conservation of mass, momentum and energy in a realistic external
environment. The formulation of the conservation laws is identical in the two
treatments.  The principal differences in the assumptions are as follows.
\begin{enumerate}
\item The analysis of LB02b explicitly assumed that there are no variations in
  physical parameters across the jets, as in our treatment of the outer
  region. We split the flaring region into laminar jet and shear layer
  components.
\item The jets in LB02b's analysis are assumed to come into approximate pressure
  equilibrium with their surroundings only after they recollimate. This then
  requires that they are over-pressured at the start of the flaring region. In
  contrast, we assume that the jets are everywhere in pressure equilibrium with
  the external medium. In this picture, the initial expansion is caused by
  transfer of momentum from the laminar core to the shear layer rather than a
  pressure-driven expansion.
\item The models are constrained in slightly different ways.  Both specify the
  radius of the jet as a function of distance from the nucleus. In LB02b, the
  velocity is given everywhere and the best average match to pressure
  equilibrium is found for the outer region. In the present model, velocities
  are specified only in the flaring region, but pressure equilibrium is enforced
  along the entire length of the jet.
\item In the solutions preferred by LB02b, momentum flux = $\Phi/c$
  initially. This is required for the jets to decelerate from
  highly-relativistic velocities on parsec scales, as in unified models of BL
  Lac objects and FR\,I radio galaxies.  It is not an explicit constraint in
  the present models, where the momentum flux is relatively higher
  (corresponding to the solutions in section 3.3.6 of LB02b).
\item We use power-law, isothermal approximations for the external density and
  pressure distributions, whereas LB02b use a double-beta-model with varying
  temperature.  The differences are minor (Fig.~\ref{density}). 
\end{enumerate}
LB02b discussed the effects of varying the assumptions of their analysis. This
led to a spread of values around those for their {\em reference model} which we
quote here.  Table~\ref{tab:compare} compares values of key parameters for our
model jet and that from LB02b's reference model at the brightening point and at
12\,kpc from the nucleus.

\begin{table}
\caption{Comparison between derived parameters for 3C\,31 in this paper and
  LB02b. Following B94 and LB02b, we quote the relativistic Mach number,
  $\mathscr{M}=\gamma_{v}v/\gamma_{c_{s}}c_{s}$, where $c_s$ is the sound speed
  and $\gamma_{c_{s}} = [1-(c_s/c)^2]^{-1/2}$. \label{tab:compare}}
\begin{tabular}{lll}
\hline
Quantity & This paper & LB02b \\
\hline
Energy flux ($10^{37}$\,W) & 1.6  & 1.1 \\
(excluding rest mass) & & \\
Initial momentum flux & 7.7 & 3.7 \\
 ($10^{28}$\,kg\,m\,s$^{-2}$) &&\\
Density at brightening point  & 12 & 2.5\\
($10^{-27}$\,kg\,m$^{-3}$)&&\\
Mass flux at brightening point  &6.2 & 1.0\\
($10^{27}$kg\,yr$^{-1}$)&&\\ 
Mass flux at 12\,kpc & 47 & 32 \\
($10^{27}$\,kg\,yr$^{-1}$) &&\\
Pressure at brightening point & 1.9  & 15 \\
($10^{-11}$\,Pa) &&\\
$\mathscr{R}$ at brightening point & 13.4 (jet)   & 0.4 \\
                                   & 7.7 (layer) & 0.4 \\
Mach number at brightening point &  7.7 (jet) & 1.5 \\
                                 &  2.5 (layer) & 1.5 \\
\hline
\end{tabular}
\end{table}

The energy fluxes of the two model jets are quite similar, despite the
differences in starting assumptions.  In terms of the available energy flux
$\Phi$ (with the rest-mass component subtracted, as in Section~\ref{jetpower}
and LB02b), we find $\Phi = 1.6 \times 10^{37}$\,W, compared with $\Phi = 1.1
\times 10^{37}$\,W for LB02b. This is because the geometries of the two jets are
identical; in the outer region their velocities are very similar and they are
both close to pressure equilibrium with the surroundings. The main difference is
in the mass flux, which is a factor of 1.5 times larger at 12\,kpc from the
nucleus in the present model.
          
There is a larger difference between the initial conditions for the two models
at the brightening point. The model jet of LB02b has an initial density roughly
5 times lower than that described here, but is also overpressured: its energy
density is dominated by the internal energy of relativistic particle rather than
by bulk kinetic energy, as can be seen from the differences in the value of
$\mathscr{R}$ at the brightening point (Table~\ref{tab:compare}).  The very low
initial density in LB02b's reference model is derived from the requirement for
FR\,I jets to be able to decelerate from bulk Lorentz factors $\sim$5 on parsec
scales. If this requirement is relaxed, as in the high-momentum solutions
described in section~3.2.6 of that paper, results closer to those in the present
paper are obtained.  The entrainment rate at the beginning of the flaring region
in both models is very low and could be provided by mass input from stars
(Section~\ref{stars}). Both models require an additional source of mass at
larger distances from the nucleus, however.  

\section{Exploring parameter space for the model}
\label{discussion}

Our model uses several parameters derived from observations of 3C\,31 to
calculate the key physical properties of this object. For other FR\,I sources,
these parameters will be inappropriate and in this section, we discuss the
effects of altering them.

\subsection{Flaring region}

The parameters affecting the solution in the flaring region are the value of
$\mathscr{R}_{e}$, the polynomial coefficients for the outer boundary, the jet and
layer velocities and the gradient of the external pressure.  We have argued that
$\mathscr{R}_{e}$, which is always very large, cannot affect our solutions
significantly. The shape of the outer boundary plays an important role in
determining the buoyancy term and varies from source to source. As the shape
function has four free parameters, we will not discuss this point in detail
here\footnote{More recent models use a two-parameter form for the shape of the
flaring region \citep{cl04,canvin05,lcbh06}.}, but we note that faster expansion
of the shear layer will lead to larger values of $\mathscr{R}_{s}(x)$ and
smaller entrainment velocities. We can vary the remaining three parameters,
$\beta_{j}$, $\beta_{s}$ and $\alpha_{2}$, individually to determine their
effect on our solutions and we plot them against $\mathscr{R}_{j}$ below. The
distributions of $\mathscr{R}_{s}$, mass flux and $v_{ent}$ are closely related
to that of $\mathscr{R}_{j}$.

\begin{figure*}
\includegraphics[width=\textwidth]{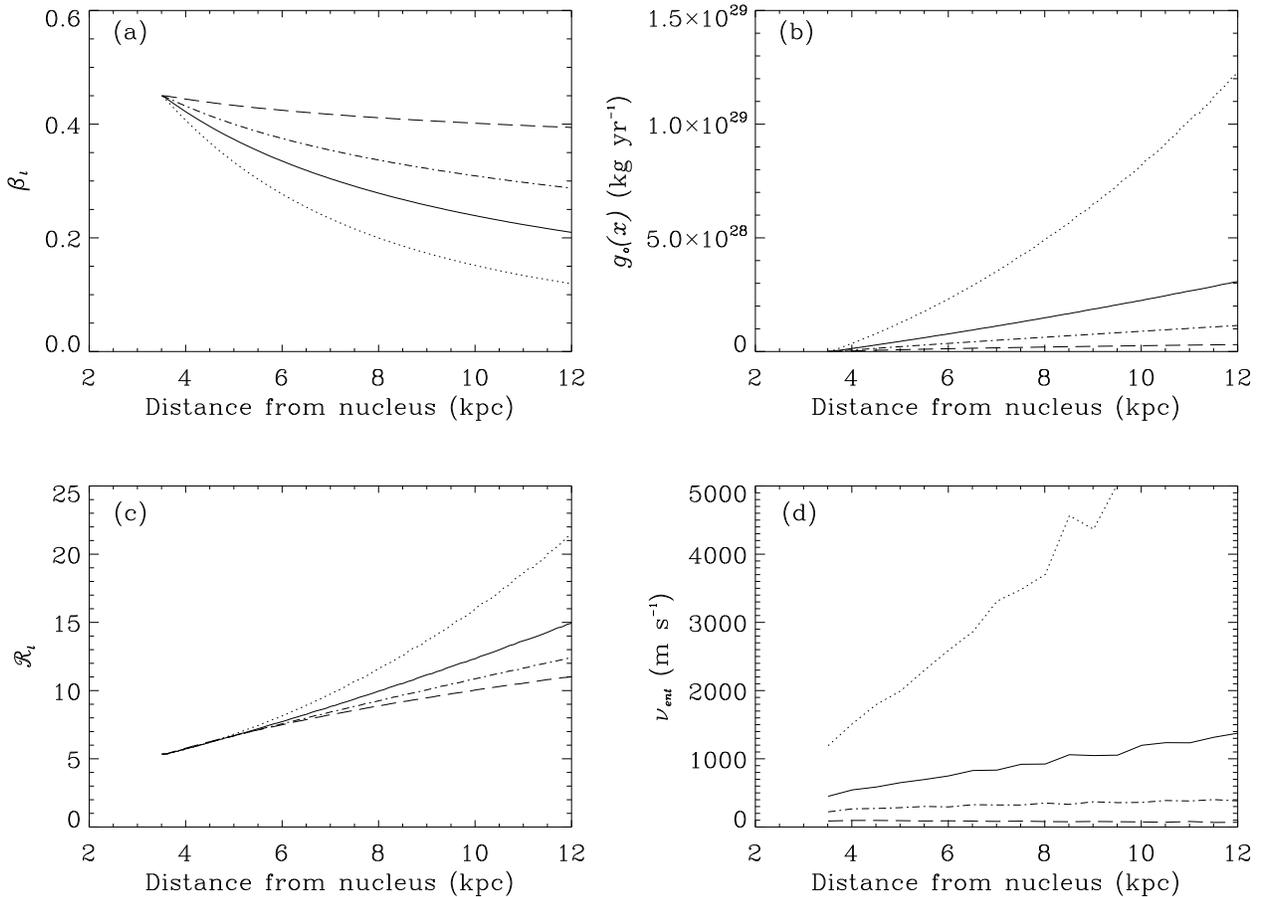}
\caption{The jet properties in the outer region for different values of
$\alpha_2$, the exponent in the external pressure  distribution. The
solid line is the value estimated for 3C\,31, $\alpha_{2}=1.1$. The dotted line,
dash dot line and dashed line are for $\alpha_{2}=0.5$, $\alpha_{2}=1.5$ and $\alpha_{2}=2$,
respectively. (a) Velocity profile, $\beta_{s}(x)$. (b) The entrainment function
$g_o(x)$. This is the entrained mass flux between the start of the outer region
($x = x_1$) and distance $x$. (c) Profile of $\mathscr{R}_{s}(x)$. (d) The
entrainment velocity perpendicular to the shear layer surface. Irregularities in
the profile are numerical artefacts.} \label{out_alpha}
\end{figure*}

Given that the laminar jet is assumed to be in pressure equilibrium with its
surroundings at the brightening point, its internal energy is determined. If
$\beta_{s}$ and the form of the pressure profile are also fixed, then the energy flux
minus the rest mass term, $\Phi$ (defined by its value at $x_1$) is also
unchanged. Since $\Phi$ is a conserved quantity, this is also true for the
laminar jet at $x_0$. A faster jet with the same internal energy must therefore
have smaller density and $R_{j}$ (Fig.~\ref{rj_betaj}).

Moreover, if we have a faster shear layer at $x_{1}$, which means that $\Phi$ is
higher, but $\beta_{j}$ remains constant, then the density of the laminar core at
$x_{0}$ must increase, since the internal energy is fixed there by the pressure
balance condition. $R_{j}$ therefore increases with $\beta_{s}$ (Fig.~\ref{rj_betal}). 
The shapes of the distributions of $g_{\rm f}(x)$, $\mathscr{R}_{s}(x)$ and
$v_{ent}(x)$ remain the same but their normalizations change if the jet or layer
velocities are varied. For a faster laminar jet or a slower shear layer,
$\mathscr{R}_{s}$ and $v_{ent}(x)$ both become smaller, indicating that the
shear layer is less dense.

The value of $\mathscr{R}_{j}$ also depends on the pressure profile, quantified
here by the exponent $\alpha_{2}$ of a power-law distribution. If the pressure
decreases more slowly with distance, then the assumption of pressure equilibrium
requires the internal energy of the layer to be higher at the end of the flaring
region, increasing the energy flux.  If the velocity of the laminar core is
fixed at the brightening point, as is its internal energy, then we need a denser
laminar jet and therefore a higher value of $\mathscr{R}_{j}$
(Fig.~\ref{rj_alpha}).

\subsection{Outer region}

For the outer region, the situation is much simpler. As $\mathscr{R}_{1}$,
$\beta_{1}$ and $r_{1}$ are determined by continuity at the boundary with the
flaring region, the only additional parameters inferred from the observations
are the half opening angle $\theta$ and the power-law exponent of the external
pressure profile, $\alpha_{2}$. Two factors influence the opening angle: the decrease of
external pressure and the expansion associated with entrainment. Of the two,
the latter is more important for 3C\,31: if we set $v_{ent}=0$ 
to remove the entrainment terms, the predicted jet opening angle is around $3^{\circ}$
(compared with the observed value of $13^\circ$), suggesting that entrainment
dominates the expansion.

Figure \ref{out_alpha} shows how the jet properties change as functions of the
exponent of the external density and pressure distributions, $\alpha_{2}$. For a
jet with a fixed opening angle, a larger value of $\alpha_{2}$ (a faster
decrease of pressure) reduces the amount of material entrained from the
environment into the jet and leads to a slower entrainment velocity. As the
buoyancy force can accelerate the material in the jet, a larger value of
$\alpha_{2}$ can also lead to a slower deceleration in the outer region. The
outer region cools due to continuous entrainment of thermal matter from the
environment into the shear layer, so $\mathscr{R}_{s}$ increases with
distance at a rate dependent on the entrainment velocity.

\begin{figure*}
\includegraphics[width=\textwidth]{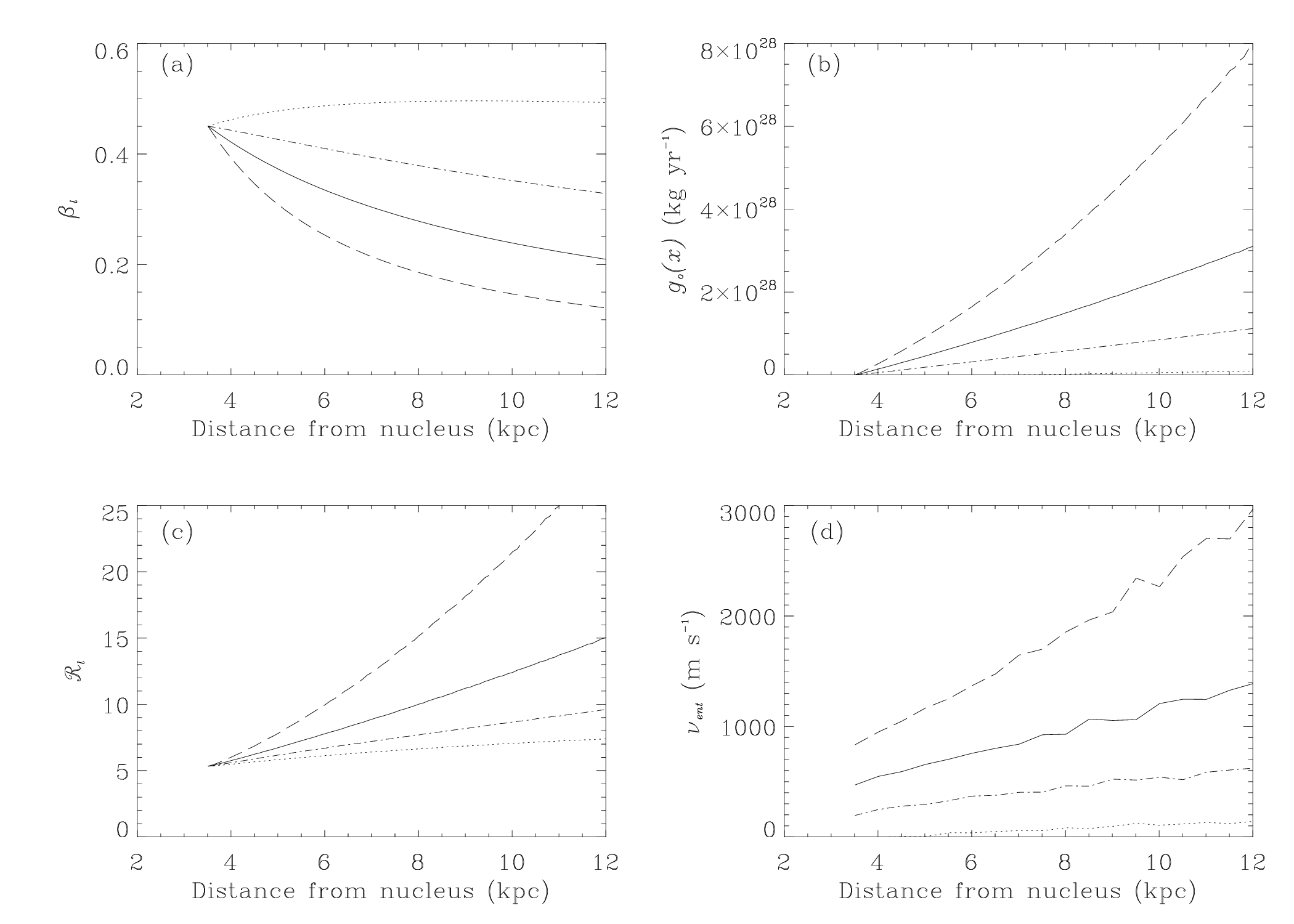}
\caption{The jet properties in the outer region for different values 
of the opening angle, $\theta$. The solid line is the default value for 3C\,31 with
$\theta=13.1^{\circ}$. The dotted line, dash dot line and dashed
line are for $\theta=3.5^{\circ},\theta=8^{\circ}$ and $\theta=20^{\circ}$
respectively.} \label{out_theta}
\end{figure*}

If we keep $\alpha_{2}=1.1$ and alter the opening angle, $\theta$, the jet
properties vary as shown in Figure \ref{out_theta}. We find that when the
opening angle is small, the jet hardly entrains any material from the
environment, and so decelerates more slowly. In extreme cases, the jet could
even be accelerated slightly by the pressure gradient.  It is interesting to
note that the other four sources which have been modelled in detail all have
outer region opening angles $<5^\circ$ \citep{cl04,canvin05,lcbh06} and show
little evidence for deceleration on these scales. Compared with 3C\,31, their
external environments are significantly less dense and it may be that
entrainment is relatively less important at large distances from the nucleus.

\section{Conclusions and Further Work}
\label{conclusion}

We have constructed an analytical mixing-layer model for jets in FR\,I radio
sources that satisfies the relativistic mass, momentum and energy conservation
laws.  FR\,I jets are observed to expand rapidly and then recollimate into
conical outflows, and we divide them into flaring and outer regions based on
this morphological distinction.  We assume that the jet is in pressure
equilibrium with its surroundings throughout both regions and divide the flaring
region into two parts: a laminar jet with very high bulk velocity, and a slower
shear layer. We prescribe the shape of the shear layer and the (constant)
velocities of the laminar jet $v_{j}$ and shear layer $v_{s}$ in the flaring
region.  We can then derive the jet power $Q$ and the ratio of rest mass energy
to non-relativistic enthalpy for the laminar jet, $R_{s}$. We calculate profiles
along the jet of the mass flux $\dot{M}(x)$, the entrainment velocity
$v_{ent}(x)$ and the ratio of rest mass energy to non-relativistic enthalpy for
the shear layer, $R_{s}(x)$.  Finally, we predict the variation of the bulk
velocity of the shear layer, $v_{s}(x)$, with distance from the nucleus in the
outer region and the radius of the laminar core $r_{s}(x)$ in the flaring region.

We have applied the model to the well-observed FR\,I radio source 3C\,31, and
find self-consistent solutions for the jet properties. In the flaring region, we
take the shape of the shear layer $r_{s}(x)$ and the bulk velocities of
$v_{j}=0.77c$ and $v_{s}=0.45c$ from fits to VLA observations (LB02a).  In the
outer region, our model predicts that the bulk velocity should decrease smoothly
to $0.22c$ at 12\,kpc, which is consistent with the values derived by LB02a. The
corresponding energy flux is $Q = 3.4 \times 10^{37}$\,W, equivalent to $\Phi =
1.6 \times 10^{37}$\,W if the rest-mass contribution is subtracted.
 
We find that $\mathscr{R}_{j}=13.4$ and that $\mathscr{R}_{s}(x)$ in the shear
layer decreases from $\approx$7.5 at the beginning of the flaring region to 6.7
and then stays almost constant until the jet recollimates. In the outer region,
$\mathscr{R}_{s}(x)$ increases from 6.7 to 15.7 at 12\,kpc, indicating that the
temperature of the material in the outer region is decreasing with distance. The
velocity of entrainment into the jet varies with distance, but has a
characteristic value of a few hundred ms$^{-1}$.

Our model gives a somewhat larger energy flux for 3C\,31 than that of LB02b, who
find $\Phi = 1.1 \times 10^{37}$\,W assuming that there are no transverse
velocity variations in the jets. The two models are quite similar in in the
outer region, but differ more significantly at the start of the flaring region:
our analysis assumes pressure equilibrium whereas LB02b require a significant
over-pressure and consequently find a lower initial density. Both models require
entrainment rates which are consistent with estimates of mass input from stars
at the base of the flaring region, but not at larger distances.

We plan to apply a slightly generalized version of our analysis to the other
FR\,I jets for which velocity models and adequate X-ray data are available
\citep{cl04,canvin05,lcbh06}.  Complex, non-axisymmetric structures are observed
at the start of the flaring regions of these jets, as they are in 3C\,31
(LB02a). It is plausible that these are shocks in the supersonic flow required
in the core, although the detailed morphology of the best-resolved example,
NGC\,315, suggests otherwise \citep{lcbh06}. Our model requires that there
should be a clear demarcation in velocity between the core and the shear layer
in FR\,I jets and predicts the shape of the former.  This can in principle be
tested using the techniques developed by LB02a, but existing observations are
limited by insufficient resolution or sensitivity in regions of rapid
deceleration close to the nucleus.\footnote{Transverse velocity gradients are
clearly detected, but they are well characterized only at larger distances from
the nucleus, where the shear layer makes up much or all of the flow in our
picture.} EVLA and e-MERLIN should be able to image the flaring regions in
detail and to resolve a core/shear-layer structure if one is present.

\section*{Acknowledgments}
YW thanks RCUK for a Dorothy Hodgkin Postgraduate Award.

\label{lastpage}

\bibliography{wy}

\end{document}